\documentclass[draft]{agujournal2019}
\usepackage{url} %this package should fix any errors with URLs in refs.
\usepackage{lineno}
\usepackage[finalnew]{trackchanges}
\usepackage{soul}
\usepackage{amssymb}
\usepackage{amsmath}
\usepackage{amsfonts}

\usepackage[ruled, vlined]{algorithm2e}
\draftfalse

\journalname{JGR: Space Physics}

\begin{document}
\sloppy

%% ------------------------------------------------------------------------ %%
%  Title
%
% (A title should be specific, informative, and brief. Use
% abbreviations only if they are defined in the abstract. Titles that
% start with general keywords then specific terms are optimized in
% searches)
%
%% ------------------------------------------------------------------------ %%

\title{Fast inverse transform sampling of non-Gaussian distribution functions in space plasmas}

%% ------------------------------------------------------------------------ %%
%
%  AUTHORS AND AFFILIATIONS
%
%% ------------------------------------------------------------------------ %%

% Authors are individuals who have significantly contributed to the
% research and preparation of the article. Group authors are allowed, if
% each author in the group is separately identified in an appendix.)

% List authors by first name or initial followed by last name and
% separated by commas. Use \affil{} to number affiliations, and
% \thanks{} for author notes.
% Additional author notes should be indicated with \thanks{} (for
% example, for current addresses).

% Example: \authors{A. B. Author\affil{1}\thanks{Current address, Antartica}, B. C. Author\affil{2,3}, and D. E.
% Author\affil{3,4}\thanks{Also funded by Monsanto.}}

\authors{Xin An\affil{1}, Anton Artemyev\affil{1}, Vassilis Angelopoulos\affil{1}, San Lu\affil{2}, Philip Pritchett\affil{3}, Viktor Decyk\affil{3}}

% \affiliation{1}{First Affiliation}
% \affiliation{2}{Second Affiliation}
% \affiliation{3}{Third Affiliation}
% \affiliation{4}{Fourth Affiliation}

\affiliation{1}{Department of Earth, Space and Planetary Sciences, University of California, Los Angeles, CA, 90095, USA}
\affiliation{2}{School of Earth and Space Sciences, University of Science and Technology of China, Hefei, 230026, China.}
\affiliation{3}{Department of Physics and Astronomy, University of California, Los Angeles, CA, 90095, USA}

%(repeat as many times as is necessary)

%% Corresponding Author:
% Corresponding author mailing address and e-mail address:

% (include name and email addresses of the corresponding author.  More
% than one corresponding author is allowed in this LaTeX file and for
% publication; but only one corresponding author is allowed in our
% editorial system.)

% Example: \correspondingauthor{First and Last Name}{email@address.edu}

\correspondingauthor{Xin An}{xinan@epss.ucla.edu}

%% Keypoints, final entry on title page.

%  List up to three key points (at least one is required)
%  Key Points summarize the main points and conclusions of the article
%  Each must be 100 characters or less with no special characters or punctuation and must be complete sentences

% Example:
% \begin{keypoints}
% \item	List up to three key points (at least one is required)
% \item	Key Points summarize the main points and conclusions of the article
% \item	Each must be 100 characters or less with no special characters or punctuation and must be complete sentences
% \end{keypoints}

\begin{keypoints}
\item New computational tool for fast sampling of arbitrary particle distribution functions is presented.
\item \change{Chebsampling has a more consistent performance on different distributions than classical rejection sampling.}{Chebyshev polynomial interpolation allows to approximate grid-based distributions and accelerates the solution of inversion problem.}
\item We illustrate the use of Chebsampling via sampling non-Harris current sheets and non-Maxwellian velocity distributions.
\end{keypoints}

%% ------------------------------------------------------------------------ %%
%
%  ABSTRACT and PLAIN LANGUAGE SUMMARY
%
% A good Abstract will begin with a short description of the problem
% being addressed, briefly describe the new data or analyses, then
% briefly states the main conclusion(s) and how they are supported and
% uncertainties.

% The Plain Language Summary should be written for a broad audience,
% including journalists and the science-interested public, that will not have 
% a background in your field.
%
% A Plain Language Summary is required in GRL, JGR: Planets, JGR: Biogeosciences,
% JGR: Oceans, G-Cubed, Reviews of Geophysics, and JAMES.
% see http://sharingscience.agu.org/creating-plain-language-summary/)
%
%% ------------------------------------------------------------------------ %%

%% \begin{abstract} starts the second page

\begin{abstract}
%[ enter your Abstract here ]
\add{Non-Gaussian distributions are commonly observed in collisionless space plasmas. Generating samples from non-Gaussian distributions is critical for the initialization of particle-in-cell simulations that investigate their driven and undriven dynamics.} \change{\texttt{Chebsampling}, a computationally efficient, robust tool to sample general distribution functions in one and two dimensions was developed.}{To this end, we report a computationally efficient, robust tool, \texttt{Chebsampling}, to sample general distribution functions in one and two dimensions.} This tool is based on inverse transform sampling with function approximation by Chebyshev polynomials. We demonstrate practical uses of \texttt{Chebsampling} through sampling typical distribution functions in space plasmas.
\end{abstract}

%\section*{Plain Language Summary}
%[ enter your Plain Language Summary here or delete this section]

%% ------------------------------------------------------------------------ %%
%
%  TEXT
%
%% ------------------------------------------------------------------------ %%

%%% Suggested section heads:
% \section{Introduction}
%
% The main text should start with an introduction. Except for short
% manuscripts (such as comments and replies), the text should be divided
% into sections, each with its own heading.

% Headings should be sentence fragments and do not begin with a
% lowercase letter or number. Examples of good headings are:

% \section{Materials and Methods}
% Here is text on Materials and Methods.
%
% \subsection{A descriptive heading about methods}
% More about Methods.
%
% \section{Data} (Or section title might be a descriptive heading about data)
%
% \section{Results} (Or section title might be a descriptive heading about the
% results)
%
% \section{Conclusions}

%\section{= enter section title =}
%Text here ===>>>
\section{Introduction}
Non-Gaussian distribution functions are commonly observed in space plasma systems, in which the extremely low frequency of particle collisions allows velocity distributions quite different from \change{the Maxwell (Gaussian) solutions}{the equilibrium solutions (Maxwellians or isotropic Gaussians)} of the Boltzmann equation \cite{bookVedenyapin11}. Except for planetary and solar atmospheres, the entire heliosphere, a region filled with plasma of solar origin, can usually be considered as a weakly collisional medium where charged particle velocity distributions may \change{arbitrarily}{significantly} deviate from the Gaussian distribution. Such velocity distributions include multicomponent distributions consisting of localized peaks in 6D phase space of velocities and coordinates, power-law distributions in which particles have a significant probability of achieving a velocity very different from the mean velocity, and distributions resulting from collisionless relaxation of plasma instabilities. \change{Although many of these distributions may be obtained in numerical simulations, this approach is computationally expensive and quite unstable. Thus, we need an approach that allows us to set any required non-Gaussian distribution as an initial condition for further investigation of its driven and undriven dynamics. For our application, we need an algorithm to sample arbitrary, non-Gaussian distributions in either configuration or velocity space to load particles in particle-in-cell simulations.}{Sampling from these non-Gaussian distributions in either configuration or velocity space is often used to load particles in kinetic simulations for further investigation of their driven and undriven dynamics. This problem in space plasmas calls for a flexible, fast sampling algorithm that can handle any non-Gaussian distributions.}

More broadly, generating pseudo-random samples from a prescribed distribution is a procedure important to computational plasma physics as well as other branches of computational physics. Particle-in-cell simulations, Monte Carlo simulations, molecular dynamics simulations, and gravitational simulations, for example, all use certain sampling algorithms to initialize various distribution functions. One of these algorithms, inverse transform sampling, is a simple method of generating samples $X$ from any probability distribution function (PDF) by inverting its cumulative distribution function (CDF) as $F_X^{-1}(U)$, where $U$ is uniformly distributed on $[0, 1]$ and $F_X$ denotes the CDF. \change{Such sampling has been thought to be unappealing}{The application of inverse transform sampling is limited in practice}, however, because it requires either a closed form of $F_X^{-1}$ or a complete approximation to $F_X$ regardless of the desired sample size, it does not generalize to multiple dimensions, and it is less efficient than other approaches \cite{gentle2003random,wilks2011statistical,givens2012computational}. Instead, rejection sampling \cite<e.g.,>[]{gentle2003random} is usually used for low-dimensional distributions, and the Metropolis-Hastings algorithm \cite{metropolis1953equation} is used for high-dimensional distributions.

The development of Chebyshev technology \cite<e.g.,>[]{trefethen2019approximation,driscoll2014chebfun} (see \url{chebfun.org}) has enabled a complete approximation to a smooth function using Chebyshev polynomials. Furthermore, approximation by Chebyshev polynomials to an analytic function converges geometrically fast \cite<see Chapter 8 in>[]{trefethen2019approximation}. For this reason, in inverse transform sampling the CDF can be well approximated to a high precision by the Chebyshev projection and can be evaluated efficiently. The use of Chebyshev projection in sampling has been only recently explored by Olver and Townsend \cite{olver2013fast}, who showed that inverse transform sampling with Chebyshev polynomial approximation is computationally efficient and robust in one dimension. They extended the approach to two dimensions but required that the CDF be well approximated by a low-rank function. 

Here we \change{develop}{report} a numerical tool to apply inverse transform sampling with Chebyshev polynomial approximation to distribution functions in one and two dimensions. In Section \ref{sec:inverse-transform-sampling-with-chebyshev}, we describe the algorithm and the implementation of our numerical tool \texttt{Chebsampling}. In Section \ref{sec-numerical-examples}, we demonstrate the accuracy and efficiency of our algorithm by sampling representative distribution functions (in either the configuration space or in the velocity space) in space plasmas. In Section \ref{sec:summary-and-discussion}, we summarize the results and discuss the pros and cons of our method.

%\section{Inverse transform sampling with Chebyshev polynomial approximation\label{sec:inverse-transform-sampling-with-chebyshev}}
\section{Methodology\label{sec:inverse-transform-sampling-with-chebyshev}}
\subsection{Inverse transform sampling}
We briefly recap the inverse transform sampling method with one and two variables. In one dimension (1D), let $f(x)$ be a PDF defined on the interval $[a, b]$. Its CDF $F_X(x)$ is a strictly increasing function. To generate $N$ samples $x_1, x_2, \cdots, x_N$ that are distributed according to $f(x)$, we invert the corresponding CDF, i.e.,
\begin{linenomath*}
\begin{equation}\label{eq:1d-inv}
	x_j = F_X^{-1} (u_j) \hspace{10pt} (j = 1, 2, \cdots, N),
\end{equation}
\end{linenomath*}
where $u_j$ is uniform on $[0, 1]$. This is inverse transform sampling. In practice, we find $x_j$ by finding the root of $F_X (x_j) = u_j$, because the inverse transform $F_X^{-1}$ often cannot be easily obtained. Thus, generating $N$ samples requires solving $N$ root-finding problems.

In two dimensions (2D), let $f(x, y)$ be a joint PDF defined on the rectangular domain $[a, b] \times [c, d]$. This joint distribution may be written as
\begin{linenomath*}
\begin{equation}
	f(x, y) = f_Y(y) \cdot f_{X \vert Y}(x \vert y),
\end{equation}
\end{linenomath*}
where $f_Y$ is the marginal distribution in the $y$ direction, and $f_{X \vert Y}$ is the conditional distribution in the $x$ direction for a given value of $y$. We do not require that $f(x, y)$ be approximated by a low-rank function as in Ref.~\cite{olver2013fast}, because this approximation is not always valid in our applications. Let $F_Y$ and $F_{X \vert Y}$ be the CDFs of $f_Y$ and $f_{X \vert Y}$, respectively. First, $N_y$ samples $y_1, y_2, \cdots, y_{N_y}$ are generated by solving the root-finding problem
\begin{linenomath*}
\begin{equation}\label{eq:inverse-trans-marginal}
	F_Y(y_k) =u_k,\,\,\, (k = 1, 2, \cdots, N_y),
\end{equation}
\end{linenomath*}
where $u_k$ is uniform on $[0, 1]$. Second, for each $y_k$ in Equation \eqref{eq:inverse-trans-marginal}, $N_x$ samples $x_{1 k}, x_{2 k}, \cdots, x_{N_x k}$ are generated by finding the root for
\begin{linenomath*}
\begin{equation}\label{eq:inverse-trans-conditional}
	F_{X \vert Y}(x_{j k} \vert y_k) = u_j, \hspace{10pt} (j = 1, 2, \cdots, N_x),
\end{equation}
\end{linenomath*}
where $u_k$ is uniform on $[0, 1]$. Thus, sampling of a joint 2D PDF is reduced to sampling of two 1D PDFs. As indicated by Equations \eqref{eq:inverse-trans-marginal} and \eqref{eq:inverse-trans-conditional}, generation of $N_x \cdot N_y$ samples requires solving $(N_x + 1) \cdot N_y$ root-finding problems. In the special case of the separable distribution function (i.e., $f(x, y) = f_Y(y) \cdot f_X(x)$), only $N_x + N_y$ root-finding problems need to be solved to generate $N_x \cdot N_y$ samples.

\subsection{Chebyshev polynomial approximation}
The efficiency of inverse transform sampling depends on the computational cost of root finding, so we adopt the bisection method for root finding. Because the CDFs increase monotonically, this method is guaranteed to converge to a high precision \cite{burden2015numerical}. We should note that it is possible to speed up the root finding further by using a hybrid bisection combined with a Newton method (since calculating derivatives with Chebyshev polynomials is fast), but it is not necessary to do so in our application, and the performance of the bisection method is acceptable. Most of the computing time in root finding is spent on evaluation of the functions (i.e., CDFs). Fortunately, most of these functions can be accurately approximated by Chebyshev polynomials, and there are well-developed fast algorithms to evaluate them. Below we describe representation of a function by Chebyshev polynomials and rapid evaluation of this function at an arbitrary point in the domain.

Chebyshev polynomials are defined on the interval $[-1, 1]$ to which other interval $[a, b]$ can be scaled. We consider the Chebyshev points
\begin{linenomath*}
\begin{equation}\label{eq-chebpts-1}
	x_k = \cos\left(\frac{k \pi}{n}\right) \hspace{20pt} (k = 0, 1, \cdots, n),
\end{equation}
\end{linenomath*}
which are extrema of the $n$th Chebyshev polynomial $T_n(x) = \cos\left(n \cdot \arccos x\right)$. The Chebyshev points are clustered near the two ends of the interval, $-1$ and $1$. Unlike polynomial interpolation at equispaced points \cite<see chap.~13 in Refs.>[]{trefethen2019approximation,platte2011impossibility}, which is associated with a well-known numerical instability (the Runge phenomenon), polynomial interpolation at the Chebyshev points is numerically stable. The Chebyshev polynomials $T_0(x)$, $T_1(x)$, $\cdots$, $T_n(x)$ on these points are orthogonal to each other \cite<see sec.~4.6.1 in Ref.>[]{mason2002chebyshev}, i.e.,
\begin{linenomath*}
\begin{equation}\label{eq-discrete-orthogonatlity}
	\sideset{}{''}\sum_{k = 0}^{n} T_i(x_k) T_j(x_k) = 
	\begin{cases}
		0,    & (0 \leqslant i, j \leqslant n;\, i \neq j),\\
		\frac{n}{2},    & (0 < i = j < n),\\
		n,    & (i = j = 0 \text{ or } n),
	\end{cases}
\end{equation}
\end{linenomath*}
where the double dash in $\sideset{}{''}\sum$ denotes the first and last terms in the sum are to be halved. This discrete orthogonality property leads us to a very efficient interpolation formula.

We approximate $f$ by the $n$th degree polynomial
\begin{linenomath*}
\begin{equation}\label{eq-cheb-interp}
	p_n(x) = \sideset{}{''}\sum_{k=0}^{n} c_k T_k(x),
\end{equation}
\end{linenomath*}
which interpolates $f$ at the Chebyshev points, i.e., $p_n(x_j) = f(x_j)$ with $x_j = \cos(j \pi / n)$. The interpolation coefficients $c_k$ are given by
\begin{linenomath*}
\begin{equation}\label{eq-discrete-chebyshev-transform}
	c_k = \frac{2}{n} \sideset{}{''}\sum_{j=0}^{n} f(x_j) T_k(x_j) \hspace{20pt} (k = 0, 1, \cdots, n).
\end{equation}
\end{linenomath*}
The evaluation of $c_k$ can be done in $\mathcal{O}\left(n \log n\right)$ operations by using the Fast Fourier Transform (FFT), which is detailed in \ref{appendix-interpolation-coeff}.

To determine the degree $n$ of the polynomial that is sufficient to approximate $f$, we adopt an adaptive procedure introduced in the Chebfun software system \cite{driscoll2014chebfun}. In this procedure, we progressively select $n$ to be $2^4 = 16$, $2^5 = 32$, $2^6 = 64$ and so on. For a given $n$, the $f$ data at the $n + 1$ Chebyshev points is converted to $n + 1$ Chebyshev coefficients. If the tail of these coefficients falls below a relative level of prescribed precision, then the Chebyshev points are judged to be fine enough. We truncate the tail and keep only the non-negligible terms. The complex engineering details of truncating a Chebyshev series are given by \remove{Ref.~}\citeA{aurentz2017chopping} (see the function ``standardChop'' in Chebfun).

\begin{figure}[tphb]
	\centering
	\includegraphics[width=\textwidth]{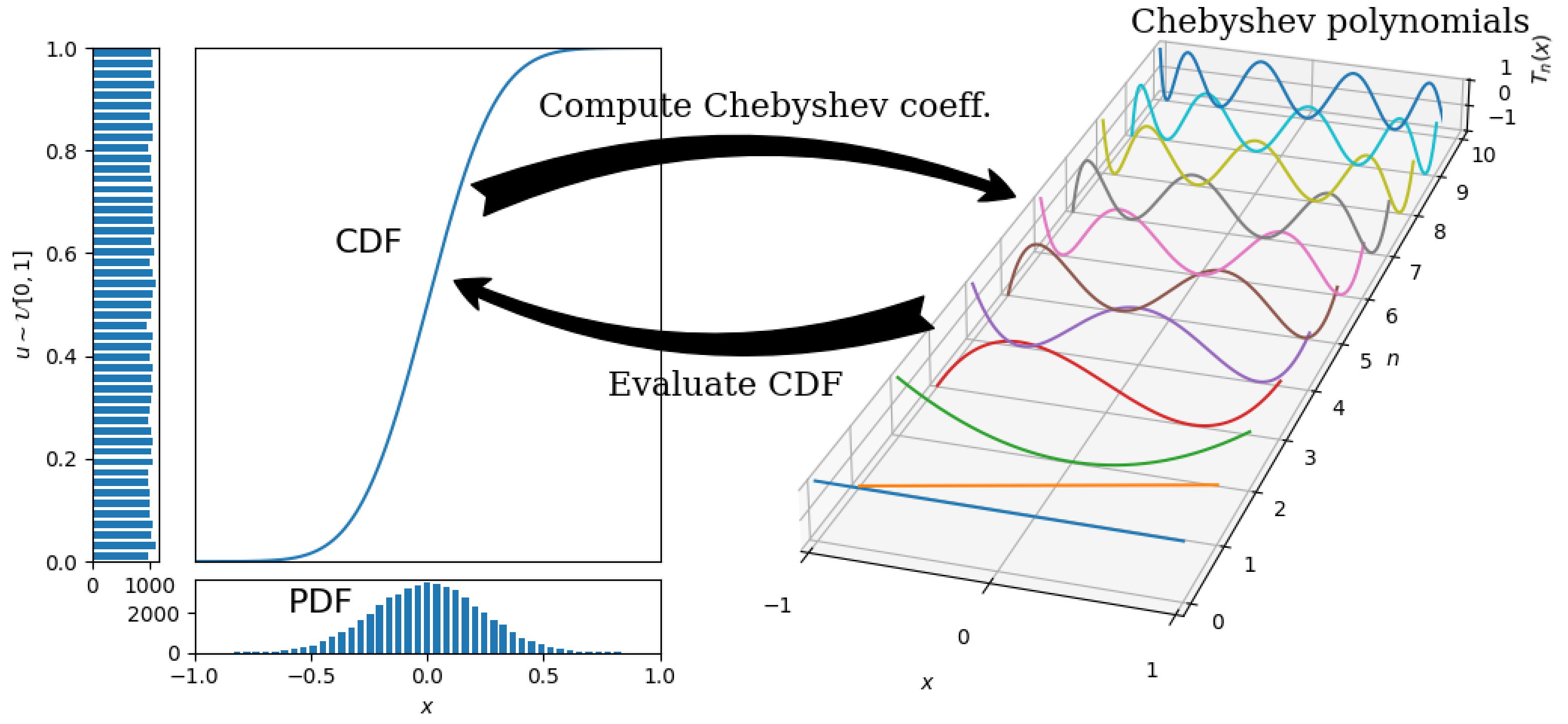}
	\caption{\label{fig:sketch}\add{Conceptual sketch for the fast inverse transform sampling with function approximation by Chebyshev polynomials. Panels on the left of the arrows represent the inversion probelm $F_X(x) = u$, where $u$ is uniformly distributed in the interval $[0, 1]$ and $F_X$ is the CDF. Panels on the right of the arrows represent the Chebyshev polynomials, which are used as the basis functions to interpolate and evaluate the CDF.}}
\end{figure}

Once the Chebyshev coefficients $c_k$ have been obtained, the original data $f$ can be discarded. These Chebyshev coefficients are then repetitively used to efficiently evaluate $f$ (also $\int f \, \mathrm{d}x$ and $\mathrm{d} f / \mathrm{d}x$) for arbitrary points in the domain. One way of achieving this is to use the Clenshaw algorithm \cite{clenshaw1955note} (details of this algorithm are described in \ref{appendix-clenshaw}). \add{To better visualize our approach desribed by Equations} \eqref{eq:1d-inv}--\eqref{eq-discrete-chebyshev-transform}, \add{we sketch the main idea of the fast inverse transform sampling with function approximation by Chebyshev polynomials in Figure} \ref{fig:sketch}.

\subsection{Implementation}
With the above considerations, we implemented \texttt{Chebsampling} in Fortran 90 with parallelization using MPI. The logical flows of 1D and 2D inverse transform sampling programs are summarized in Algorithms \ref{inv_sampling_1d} and \ref{inv_sampling_2d}, respectively. Notably, our input PDF data are defined on grid, which is more flexible when an analytical expression of the input PDF is not available. For the 2D joint PDF, we apply the 1D sampling algorithm repetitively to generate samples from marginal and conditional distribution functions. As demonstrated in Section \ref{sec-numerical-examples}, inverse transform sampling using the Chebyshev polynomial approximation is very efficient.

To generate a large number of samples, we parallelize the 2D inverse transform sampling algorithm. First, $N_{py}$ samples of $y$ are drawn from the marginal distribution function $f_Y(y)$, which is executed on all processors. Second, the tasks of sampling the conditional distribution function $f_{X \vert Y}(x \vert y)$ are evenly divided among processors based on the $y$ samples, such that the load is balanced on each processor. The $x$ samples drawn from the conditional distribution function are stored in local memory. This parallelization scheme yields nearly ideal scaling of the computational cost against the number of processors (see performance tests in Section \ref{sec-numerical-examples}).

\begin{algorithm}[H]
	\SetAlgoLined
	\KwIn{PDF data $f(x_j)$ defined on a 1D uniform grid ($j = 1, 2, \cdots, N_g$) and the desired number of samples $N_{\mathrm{samples}}$.}
	\KwOut{Samples $x_m$ ($m = 1, 2, \cdots, N_{\mathrm{samples}}$) distributed according to $f(x)$.}
	\begin{itemize}
		\item Calculate the cumulative sum of $f(x_j)$ using the recursive relation $F(x_j) = F(x_{j-1}) + \left(f(x_{j-1}) + f(x_{j})\right) / 2$ where $F(x_1) = 0$ and $j = 2, 3, \cdots, N_g$\;
		\item Normalize the cumulative sum as $F(x_j) = F(x_j) / F(x_{N_g})$\;
		\item \change{Construct the Chebyshev coefficients $c_l$ ($l = 0, 1, \cdots, N_{\mathrm{cutoff}}$) from data $F(x_j)$\;}{Progressively select $n = 16, 32, 64, \cdots$ as in a loop, compute the Chebyshev coefficients $c_l$ ($l = 0, 1, \cdots, n$) using Equation~(8), and exit the loop if the tail of these coefficients falls below a relative level of prescribed precision;}
		\item Generate samples $x_m$ by solving the root-finding problem $F(x_m) = u_m$ with the bisection method, where $F(x) = \sum_{l = 0}^{N_\mathrm{cutoff}} c_l T_l(x)$, $u_m = (m - 0.5) / N_\mathrm{samples}$, and $m = 1, 2, \cdots, N_{\mathrm{samples}}$.
	\end{itemize}
	\caption{1D inverse transform sampling using the Chebyshev polynomial approximation.\label{inv_sampling_1d}}
\end{algorithm}

\begin{algorithm}[H]
	\SetAlgoLined
	\KwIn{PDF data $f(x_j, y_k)$ defined on a 2D uniform grid ($j = 1, 2, \cdots, N_{gx}$; $k = 1, 2, \cdots, N_{gy}$); the desired number of samples in $x$ direction $N_{px}$; the desired number of samples in $y$ direction $N_{py}$.}
	\KwOut{Samples $(x_{mn}, y_n)$ ($m = 1, 2, \cdots, N_{px}$; $n = 1, 2, \cdots, N_{py}$) distributed according to $f(x, y)$.}
	\begin{itemize}
		\item Calculate the marginal distribution function $f_Y(y_k) = \sideset{}{''}\sum_{j = 1}^{N_{gx}} f(x_j, y_k)$\;
		\item Draw samples $y_n$ ($n = 1, 2, \cdots, N_{py}$) from the marginal distribution $f_Y(y)$ by performing 1D inverse transform sampling using the Chebyshev polynomial approximation\;
		\item For each sample $y_n$ ($n = 1, 2, \cdots, N_{py}$):
		\begin{itemize}
			\item Construct the conditional distribution function $f_{X \vert Y}(x_j \vert y_n)$ by interpolating $f(x_j, y_k)$ ($j = 1, 2, \cdots, N_{gx}$; $k = 1, 2, \cdots, N_{gy}$) into sampled locations $y_n$\;
			\item Draw samples $x_{mn}$ ($m = 1, 2, \cdots, N_{px}$) from the conditional distribution $f_{X \vert Y}(x \vert y)$ by performing 1D inverse transform sampling with the Chebyshev polynomial approximation.
		\end{itemize}
	\end{itemize}
	\caption{2D inverse transform sampling using the Chebyshev polynomial approximation.\label{inv_sampling_2d}}
\end{algorithm}

\section{Numerical examples\label{sec-numerical-examples}}
Below we illustrate the performance and accuracy of our algorithm by applying it to representative distribution functions in space plasmas.

\subsection{2D Maxwellian current sheets}
We first consider the density distribution relevant to a very important plasma equilibrium, the 2D current sheet, which is believed to be formed in the solar corona and has been commonly observed in planetary magnetospheres. In their seminal paper, Lembege and Pellat \cite{lembege1982stability} constructed a two-dimensional current sheet at equilibrium that resembles the planetary magnetotail configuration. In this model, the magnetic field lines in the $x$-$z$ plane are described by the vector potential $A_y (\varepsilon x, z) \mathbf{e}_y$, where $\vert \varepsilon \vert \ll 1$ indicates weak dependence of $A_y$ on $x$. The vector potential is determined by Ampere's law
\begin{linenomath*}
\begin{equation}
	\frac{\partial^2}{\partial z^2} A_y = - 4 \pi \sum_{\alpha} q_\alpha n_0 \frac{v_{D \alpha}}{c} \exp\left( -\frac{q_\alpha \varphi}{T_{\alpha 0}} + \frac{v_{D\alpha} q_\alpha A_y}{c T_{\alpha 0}} \right) \label{eq-Ay-Lembege-Pellat},
\end{equation}
\end{linenomath*}
where $\varphi (x, z)$ is the electrostatic potential, $n_0$ is the reference density, $v_{D \alpha}$ is the drift velocity, $T_{\alpha 0}$ is the temperature of current sheet particles, $q_{\alpha}$ is the charge, and $c$ is the speed of light. The subscript $\alpha = e, i$ represents electrons and ions, respectively. Note that $\partial^2 A_y / \partial x^2$ is omitted in Equation \eqref{eq-Ay-Lembege-Pellat}, and thus the equation is precise to order $\varepsilon$. The current density in Equation \eqref{eq-Ay-Lembege-Pellat} is derived by integrating the Boltzmann-type distribution in velocity space. The electrostatic potential $\varphi$ is determined by the quasi-neutrality condition
\begin{linenomath*}
\begin{equation}\label{eq-quasi-neutrality-2}
	\sum_{\alpha} q_\alpha n_0 \exp\left( -\frac{q_\alpha \varphi}{T_{\alpha 0}} + \frac{v_{D \alpha} q_\alpha A_y}{c T_{\alpha 0}} \right) + q_\alpha n_b \exp\left( -\frac{q_\alpha \varphi}{T_{\alpha b}} \right) = 0.
\end{equation}
\end{linenomath*}
Here two populations, the current sheet population (i.e., the current-carrying one) and the background population (i.e., the non-current-carrying one), are represented by the first and the second terms, respectively. In this example, we solve Equations \eqref{eq-Ay-Lembege-Pellat} and \eqref{eq-quasi-neutrality-2} in the rectangular domain $[-L_z / 2 \leqslant z \leqslant L_z / 2] \times [-L_x \leqslant x \leqslant 0]$ with the boundary condition $A_y \big\vert_{z = 0} = \varepsilon B_0 x,\,\, \partial A_y / \partial z \big\vert_{z = 0} = 0$. Here $B_0$ refers to the asymptotic magnetic field at $z \to \pm \infty$, and $\varepsilon B_0$ gives the $z$ component of the magnetic field at $z = 0$. An analytical solution of $A_y$ and $\varphi$ is not available except for the particular choice of parameters, i.e., $v_{Di} / T_{i0} = - v_{De} / T_{e0}$. To handle more general scenarios, we solve Equations \eqref{eq-Ay-Lembege-Pellat} and \eqref{eq-quasi-neutrality-2} numerically and obtain $A_y$ and $\varphi$ on grid.

Table \ref{tab:lembege-pellat-parameters} shows the two sets of parameters that are used as examples below. The first set of parameters satisfies $v_{Di} / T_{i0} = - v_{De} / T_{e0}$ so the electrostatic potential is zero everywhere in the domain (a nonpolarized current sheet). The second set of parameters has the relation $ \vert v_{Di} / T_{i0} \vert < \vert v_{De} / T_{e0} \vert$, and thus gives a nonzero electric field (a polarized current sheet). This plasma equilibrium is used as an initial condition for numerical simulations helpful in solving many problems
related to plasma stability and dynamics in planetary magnetotails. Therefore, a critical task is to generate a 2D spatial distribution of plasma particles for a given numerical solution of scalar and vector potentials. For purposes of demonstration,  we apply our method to sample the density distribution of the current sheet population,
\begin{linenomath*}
\begin{equation}\label{eq:density-current-sheet-population}
	n_{\alpha 0} = n_0 \exp\left( -\frac{q_\alpha \varphi}{T_{\alpha 0}} + \frac{v_{D \alpha} q_\alpha A_y}{c T_{\alpha 0}} \right) .
\end{equation}
\end{linenomath*}

\begin{table}
	\centering
	\def\arraystretch{1.5}%  1 is the default, change whatever you need
	\begin{tabular}{c | c | c | c | c | c | c | c | c | c}
		\hline
		& $\frac{v_{Di}}{v_\mathrm{A}}$ & $\frac{v_{De}}{v_\mathrm{A}}$ & $\frac{T_{i0}}{m_i v_\mathrm{A}^2}$ & $\frac{T_{e0}}{m_i v_\mathrm{A}^2}$ & $\frac{T_{ib}}{m_i v_\mathrm{A}^2}$ & $\frac{T_{eb}}{m_i v_\mathrm{A}^2}$ & $\frac{n_b}{n_0}$ & $\frac{L_x}{d_i}$ & $\frac{L_z}{d_i}$ \\
		\hline
		Nonpolarized & $\frac{5}{12}$ & $-\frac{1}{12}$ & $\frac{5}{12}$ & $\frac{1}{12}$ & $\frac{5}{12}$ & $\frac{1}{12}$ & $0.2$ & $32$ & $16$ \\
		%\hline
		Polarized & $\frac{1}{3}$ & $-\frac{4}{3}$ & $\frac{5}{12}$ & $\frac{1}{12}$ & $\frac{5}{12}$ & $\frac{1}{12}$ & $0.2$ & $32$ & $16$ \\
		\hline
	\end{tabular}
	\caption{\label{tab:lembege-pellat-parameters}Two sets of parameters for nonpolarized and polarized Lembege-Pellat current sheets. The velocities are normalized to the Alfv\'en velocity $v_\mathrm{A} = B_0 / \sqrt{4 \pi n_0 m_i}$, the temperatures are normalized to $m_i v_\mathrm{A}^2$, the densities are normalized to $n_0$, and the length is normalized to the ion inertial length $d_i$.}
\end{table}

\begin{figure}[tphb]
	\centering
	\includegraphics[width=\textwidth]{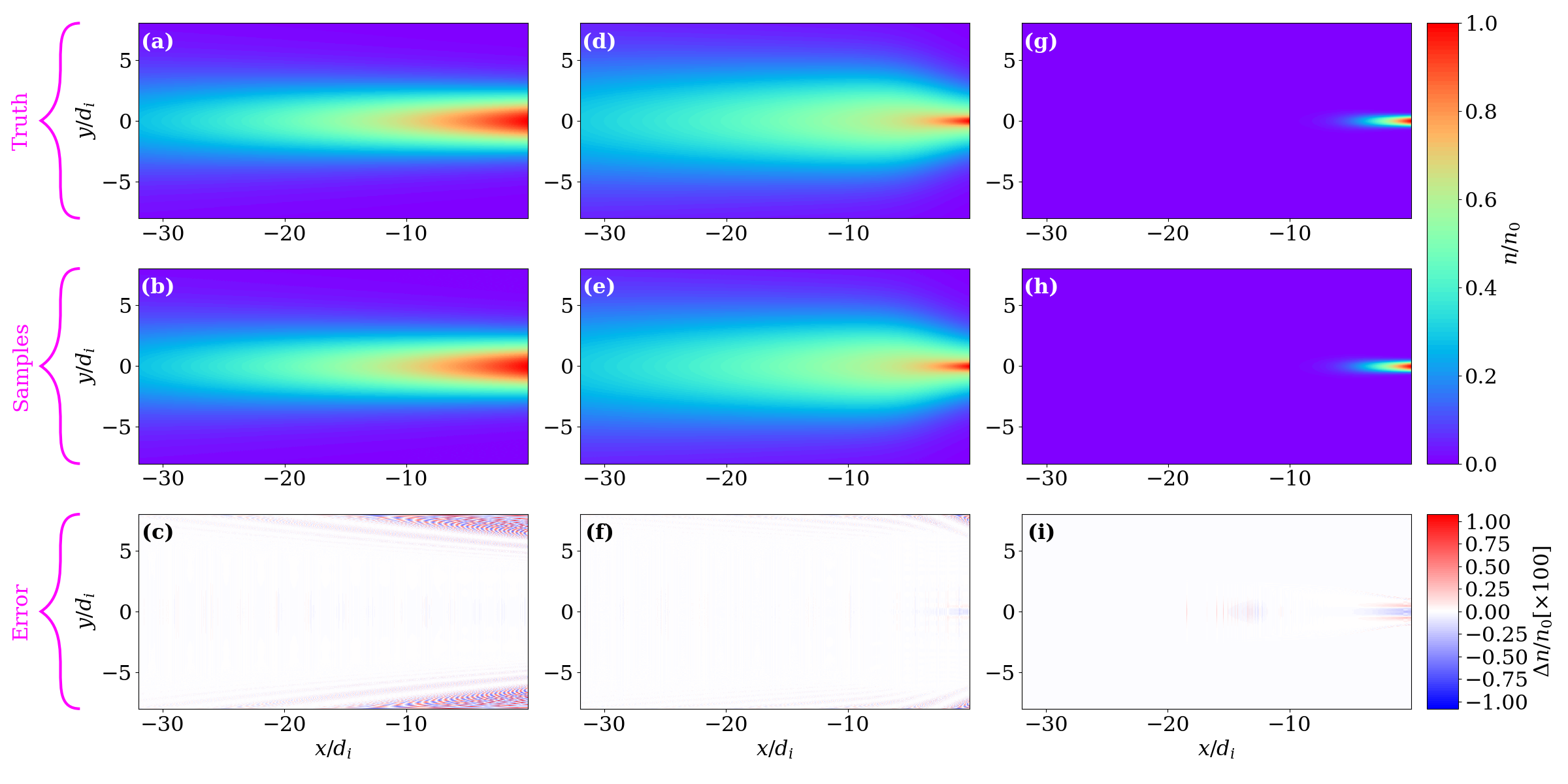}
	\caption{\label{fig:lembege-pellat-2d}Inverse transform sampling of the Lembege-Pellat current sheet. (a)-(c) The nonpolarized Lembege-Pellat current sheet set up using the first set of parameters in Table \ref{tab:lembege-pellat-parameters}. (d)-(i) The ion component [(d), (e), (f)] and the electron component [(g), (h), (i)] of the polarized Lembege-Pellat current sheet. This current sheet is obtained using the second set of parameters in Table \ref{tab:lembege-pellat-parameters}. The displayed distributions are for the current sheet population only, as shown in Equation \eqref{eq:density-current-sheet-population}. The three rows from top to bottom show the ground-truth density distributions, the sampled density distributions, and the difference between the sampled and ground-truth distributions, respectively.}
\end{figure}

\begin{figure}[tphb]
	\centering
	\includegraphics[width=0.6\textwidth]{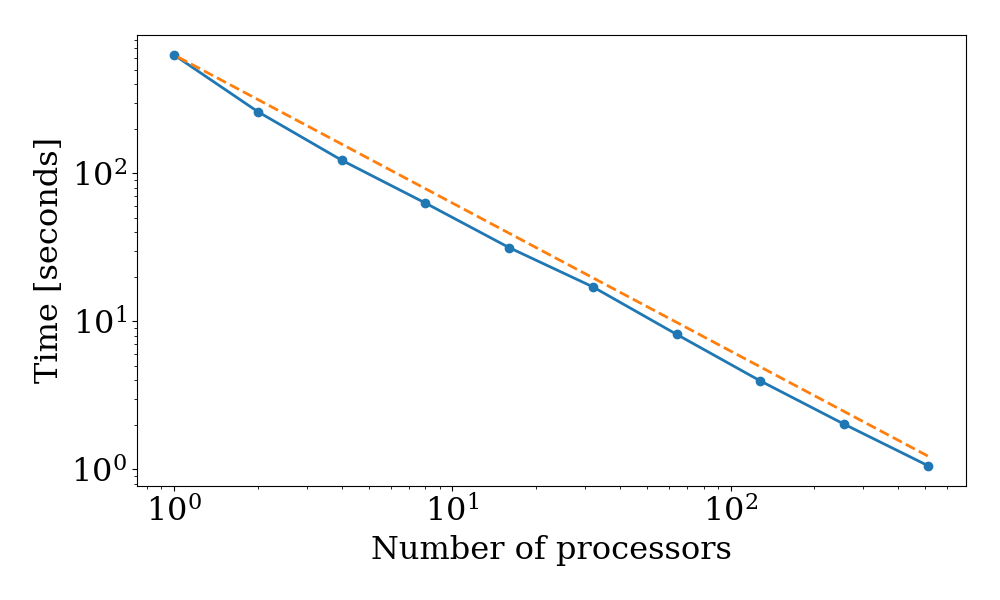}
	\caption{\label{fig:timings}\add{Strong scaling of \texttt{Chebsampling} for sampling the nonpolarized Lembege-Pellat current sheet. The $x$ and $y$ axes represent  the number of processors and the elapsed wall-clock time, respectively.} \remove{The elapsed wall-clock time versus the number of processors generating samples for the nonpolarized Lembege-Pellat current sheet.} The dashed line represents the ideal scaling.}
\end{figure}

For the first set of parameters, the density distribution of the ion current sheet is identical to that of the electron current sheet. To sample this density distribution, we use $N_{px} = 20000$ particles in the $x$ direction and $N_{pz} = 10000$ particles in the $z$ direction, which gives a total of $N_{px} \cdot N_{pz} = 2 \times 10^8$ particles. Figures \ref{fig:lembege-pellat-2d}(a) through \ref{fig:lembege-pellat-2d}(c) show the excellent agreement between the ground truth density and the sampled density. The errors [$\lesssim 1\%$; Figure \ref{fig:lembege-pellat-2d}(c)] come from the low-density region and are negligible for our application (particle-in-cell simulations). This sampling takes about $628$ seconds on a single processor. Using the parallelization scheme outlined in Section \ref{sec:inverse-transform-sampling-with-chebyshev}, we observe that the sampling takes about $1$ second on $512$ processors. As shown in Figure \ref{fig:timings}, the wall-clock time used in sampling scales ideally against the number of processors.

Similarly, we generate $2 \times 10^8$ samples for the polarized current sheet. The results are shown in Figures \ref{fig:lembege-pellat-2d}(d) through \ref{fig:lembege-pellat-2d}(i). In this case, the electron current sheet [Figure \ref{fig:lembege-pellat-2d}(g)] is embedded in the ion current sheet [Figure \ref{fig:lembege-pellat-2d}(d)]. Sampling the electron current sheet is challenging because of the steep gradient at its edge. The Chebyshev projection, which is able to capture the main characteristics of the electron current sheet, gives an accurate sampled distribution [Figures \ref{fig:lembege-pellat-2d}(h)-(i)].

In Table \ref{tab:rs-its-comparison}, we compare the performance of inverse transform sampling with rejection sampling for the distributions shown in Figure \ref{fig:lembege-pellat-2d}. For relatively fat distribution functions as in Figures \ref{fig:lembege-pellat-2d}(a) and \ref{fig:lembege-pellat-2d}(d), rejection sampling is more efficient than inverse transform sampling. For highly peaked distribution functions as in Figure \ref{fig:lembege-pellat-2d}(g), however, inverse transform sampling outperforms rejection sampling. To represent such a distribution function, inverse transform sampling must only add more Chebyshev coefficients that do not add much computational cost, whereas rejection sampling rejects a significant fraction of samples that does add much computational cost (because the ratio of the area under the distribution function to that under the rectangular hat function is small). Therefore, inverse transform sampling avoids the practical limit in rejection sampling and gives a more consistent performance across distribution functions with vastly different shapes.

\begin{table}
	\centering
	\def\arraystretch{1.5}%  1 is the default, change whatever you need
	\begin{tabular}{c | c | c}
		\hline
		& ITS [seconds] & RS [seconds] \\
		\hline
		Nonpolarized current sheet & $3.2$ & $0.75$ \\
		Polarized ion current sheet & $3.6$ & $0.6$ \\
		Polarized electron current sheet & $9.7$ & $30.6$\\
		\hline
	\end{tabular}
	\caption{\label{tab:rs-its-comparison} Performance comparison of inverse transform sampling (ITS) with rejection sampling (RS). In this comparison, $2 \times 10^6$ samples are generated using a single processor for each case. The current sheet distributions in the three rows correspond to Figures \ref{fig:lembege-pellat-2d}(a), \ref{fig:lembege-pellat-2d}(d), and \ref{fig:lembege-pellat-2d}(g), respectively.}
\end{table}

\subsection{Non-Maxwellian velocity distributions}
\begin{figure}[tphb]
	\centering
	\includegraphics[width=\textwidth]{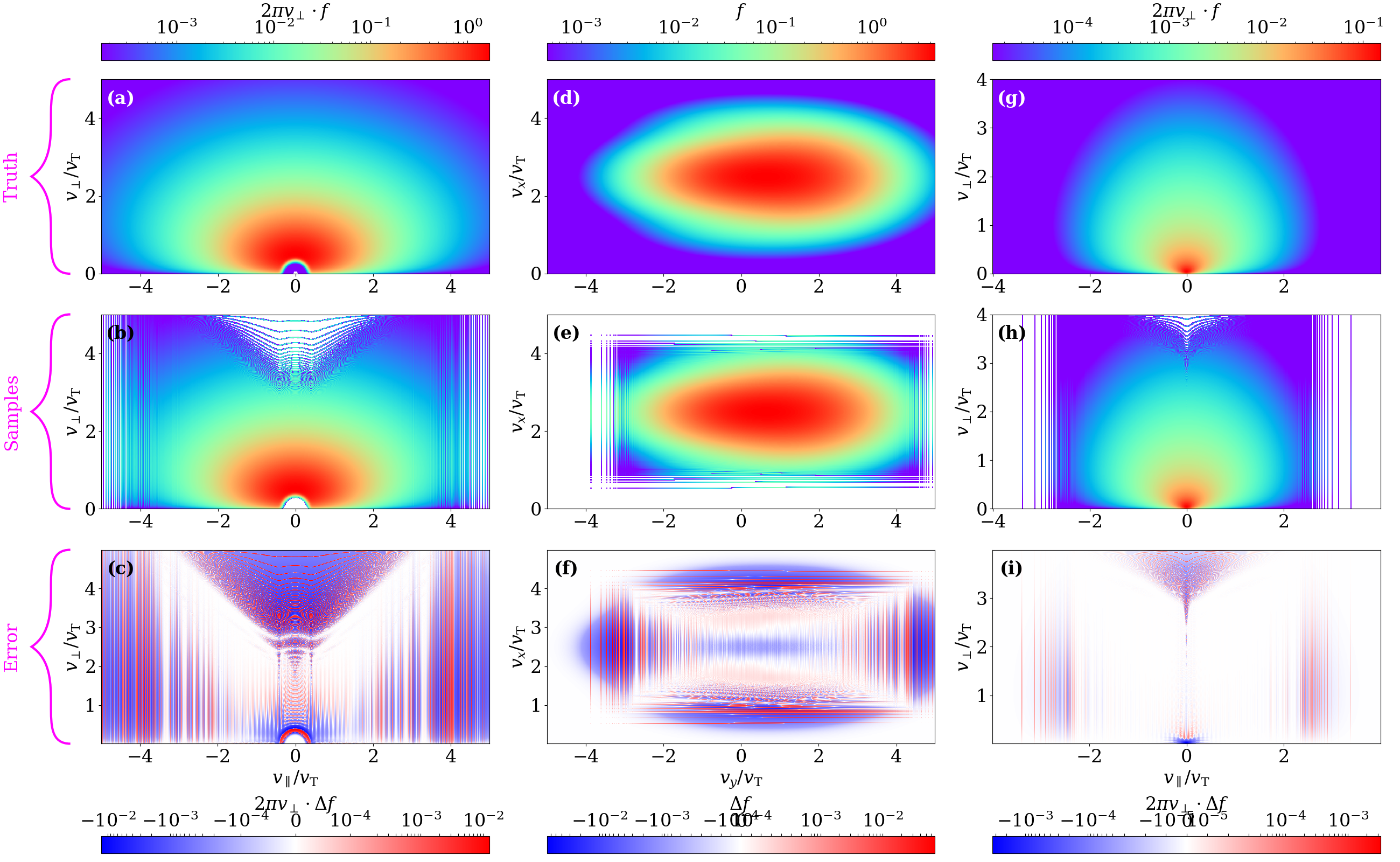}
	\caption{\label{fig:vdistr_2d}Inverse transform sampling of velocity distributions in space plasmas. (a)-(c) Halo electrons in the solar wind. (d)-(f) The electron distribution in the force-free current sheet. (g)-(i) The electron distribution in the injection regions in Earth's magnetotail. The three rows from top to bottom show the ground-truth velocity distributions, the sampled velocity distributions, and the difference between the sampled and ground-truth distributions, respectively. \add{The sampling errors appearing as lines instead of random dots are caused by setting nonrandom, evenly distributed $u_k = (k - 0.5) / N_y\,\, (\mathrm{where}\,\, k = 1, 2, \cdots, N_y)$ in Equation} \eqref{eq:inverse-trans-marginal} [\add{similarly for $u_j$ in Equation} \eqref{eq:inverse-trans-conditional}] \add{instead of random samples uniformly distributed in the interval [0, 1], which corresponds to the ``quiet start''} \cite{birdsall2018plasma}.}
\end{figure}

Furthermore, we consider three non-Maxwellian velocity distributions in the solar wind and the terrestrial magnetosphere:
\begin{enumerate}
	\item Halo electrons in the solar wind \cite{vstverak2009radial}: 
	\begin{linenomath*}
	\begin{equation}\label{eq:halo-distr}
		\begin{split}
			f (v_\perp, v_\parallel) =& \left\{ 1- \left[1 + \left(\frac{1}{2 \delta} \left(\frac{v_{\perp}^2}{v_{c \perp}^2} + \frac{v_{\parallel}^2}{v_{c \parallel}^2}\right)\right)^p\right]^{-q} \right\} \\
			&\times \left[1 + \frac{1}{2 \kappa_h - 3} \left(\frac{v_{\perp}^2}{v_{h \perp}^2} + \frac{v_{\parallel}^2}{v_{h \parallel}^2}\right)\right]^{-\kappa_h - 1}
		\end{split}
	\end{equation}
	\end{linenomath*}
	with $\kappa_h = 3$, $v_{h \parallel} = v_T = 1$, $v_{h \perp} = 1 / \sqrt{2}$, $v_{c \parallel} = 0.3$, $v_{c \perp} = 0.3$, $\delta = 0.9$, $p$ = 10 and $q = 1$;
	\item Electrons in the force-free current sheet \cite{harrison2009one}: 
	\begin{linenomath*}
	\begin{equation}\label{eq:force-free-distr}
		\begin{split}
			f(v_x, v_y, v_z) =& \exp\left(-\beta \frac{v_x^2 + v_y^2 + v_z^2}{2}\right) \\
			&\times \left[\exp\left(\beta u_y (v_y + A_y)\right) + a \cos\left(\beta u_x (v_x + A_x)\right) + b\right]
		\end{split}
	\end{equation}
	\end{linenomath*}
	with $\beta = v_{T}^{-2} = 1$, $u_x = u_y = \sqrt{2}$, $A_x = A_y = 0$, $a = 1$ and $b = 2$;
	\item Electrons in the injection regions in the Earth's magnetotail \cite{damiano2015ion,vasko2017electron,artemyev2020ionosphere}:
	\begin{linenomath*}
	\begin{equation}\label{eq:power-maxwellian}
		f (v_\perp, v_\parallel) = \left[1 + \frac{1}{\kappa} \left(\frac{v_\perp^2}{v_{c \perp}^2} + \frac{v_\parallel^2}{v_{c \parallel}^2}\right) \right]^{-\kappa - 1} \exp\left(-\frac{v_\perp^2}{2 v_{h \perp}^2} - \frac{v_\parallel^2}{2 v_{h \parallel}^2}\right)
	\end{equation}
	\end{linenomath*}
	with $\kappa = 0.2$, $v_{h \parallel} = v_T = 1$, $v_{h \perp} = \sqrt{2}$, $v_{c \perp} = \sqrt{3/800} v_{h \perp}$ and $v_{c \parallel} = \sqrt{2} v_{c \perp}$.
\end{enumerate}
The velocity distributions in Equations \eqref{eq:halo-distr} and \eqref{eq:power-maxwellian} are uniform in gyrophase, and the velocity distribution in Equation \eqref{eq:force-free-distr} obeys a Maxwellian in the $z$ direction that is separable from the $x$ and $y$ directions. Thus, these sampling problems are essentially two dimensional. Figure \ref{fig:vdistr_2d} shows the results of generating $2 \times 10^8$ samples for each of the three velocity distributions. The sampling times for these three cases are about $7$--$9$ seconds on $64$ processors. The sampled distributions capture the main trends of the original distributions. The errors are located at the high-energy tails, where the number of particles is limited. For our application in particle-in-cell simulations, such errors will not cause any problem, because the fraction of high-energy particles is very small, and thus their contribution to the charge and current deposition is small compared to the bulk of the distribution. It is noteworthy that the Chebyshev projection can fit the flat-top part of the halo electron distribution [i.e., the truncated core of the distribution with almost no electrons; see Figures \ref{fig:vdistr_2d}(a) and \ref{fig:vdistr_2d}(b)]. Because such flat-top distributions have also been found in the magnetic reconnection region \cite{asano2008electron} and the shock region \cite{wilson2019electron}, sampling them could be useful for other studies.

%\textcolor{red}{\subsection{Performance evaluation}}

\add{We list the computational details in sampling the six representative distribution functions in Table} \ref{tab:performance-info}. \add{From each distribution, $2 \times 10^8$ sampels are generated on $64$ processors. For all cases, the time it takes to compute the Chebyshev coefficients ($t_{\mathrm{cheb}}$) is about $1000$ times shorter than that of the bisection root finding ($t_{\mathrm{bisc}}$). For distributions that have steep gradients such as in the polarized current sheet, the number of Chebyshev coefficients can be large and thus it takes longer to compute those coefficients.}

\begin{table}
	\centering
	\def\arraystretch{1.5}%  1 is the default, change whatever you need
	\begin{tabular}{c | c | c | c | c | c | c}
		\hline
		& $\epsilon_{\mathrm{cheb}}$ & $N_{\mathrm{cutoff}}$ & $t_{\mathrm{cheb}}$ [ms] & $\epsilon_{\mathrm{bisc}}$ & $N_{\mathrm{bisc}}$ & $t_{\mathrm{bisc}}$ [s] \\
		\hline
		Nonpolarized current sheet & $10^{-8}$ & $7$--$38$ & $8.2$ & $10^{-14}$ & $46$--$48$ & $11.3$ \\
		Polarized ion current sheet & $10^{-8}$ & $16$--$144$  & $16$ & $10^{-14}$ & $46$--$48$  & $48.1$ \\
		Polarized electron current sheet & $10^{-8}$ & $24$--$190$ & $59$ & $10^{-14}$ & $48$ & $69.2$ \\
		Halo electrons in the solar wind & $10^{-8}$ & $14$--$360$ & $6.8$ & $10^{-14}$ & $47$--$48$ & $8.5$ \\
		Electrons in force-free current sheets & $10^{-8}$ & $24$--$28$ & $7.9$ & $10^{-14}$ & $48$ & $7.9$ \\
		Electrons in the injection fronts & $10^{-8}$ & $12$--$376$ & $4.6$ & $10^{-14}$ & $47$--$48$ & $6.4$ \\
		%\hline
		\hline
	\end{tabular}
	\caption{\label{tab:performance-info}\add{Computational details of the six representative distribution functions. $\epsilon_{\mathrm{cheb}}$ and $\epsilon_{\mathrm{bisc}}$ are the relative error controls in Chebyshev polynomial interpolation and bisection root finding, respectively. $N_{\mathrm{cutoff}}$ is the number of Chebyshev coefficients to interpolate each distribution using Chebyshev polynomials. $N_{\mathrm{bisc}}$ is the number of iteration for convergence of root finding using the bisection method. $t_{\mathrm{cheb}}$ and $t_{\mathrm{bisc}}$ are the time costs of Chebyshev polynomial interpolation and bisection root finding, respectively. Note that $N_{\mathrm{cutoff}}$ is shown as a range because it varies for the marginal and conditional distributions.}}
\end{table}

\section{Summary and discussion\label{sec:summary-and-discussion}}
We developed a novel tool, \texttt{Chebsampling}, for accurate, efficient sampling of distribution functions in one and two dimensions. It features the use of function approximation by Chebyshev polynomials, which accelerates root finding in the inverse transform sampling. \texttt{Chebsampling} is implemented on massively parallel computers and has the potential to be used for fully three-dimensional sampling in physical systems. The practical use of this tool is illustrated through typical examples in space plasmas.

Inverse transform sampling is efficient for any distribution functions that can be numerically approximated and evaluated with low cost. The distribution function can be well approximated in one dimension by Chebyshev polynomials, and the inverse sampling method is practical. The sample size in two or three dimensions is relatively small and the time cost is affordable with parallelizations. With increasing sample size, however, using the inverse transform sampling in higher dimensions is challenging\change{.}{, because one needs to perform approximately the same number of inversions as the sample size.} \add{Although function approximation in two dimensions starts to emerge} \cite{townsend2013extension}, fundamental algorithmic issues on how to numerically approximate general distribution functions with more variables remain. Once these issues have been resolved, the inverse transform sampling method will be immediately usable in higher dimensions. \add{Rejction sampling has a similar problem in higher dimensions. As the dimensions get larger, the ratio of the embedded volume to the total volume goes to zero. Thus a significant number of unwanted samples are rejected before a useful sample is obtained. In high dimensions, the Metropolis-Hastings algorithm is usually used, which is beyond the scope of our study.}  

\section{Open research}
\add{The code \texttt{Chebsampling} that has been developed in this manuscript is publicly available at} \url{https://doi.org/10.5281/zenodo.6109523}. \add{A compute capsule for reproducing the runs in this manuscript has been set up at} \url{https://codeocean.com/capsule/0988490/tree/v2}.

\appendix
\section{Computation of the Chebyshev coeffcients $c_k$ \label{appendix-interpolation-coeff}}
The evaluation of the Chebyshev coefficients $c_k$ through the use of FFT has been well established \cite<see Refs.~>[]{mason2002chebyshev,ahmed1968study,orszag1971accurate,orszag1971galerkin}. Equation \eqref{eq-discrete-chebyshev-transform} can be viewed as the discrete Chebyshev transform $f(x_k) \to c_k$. The connection to discrete Fourier transform can be seen through a change in variables
\begin{linenomath*}
\begin{equation}
	g(\theta) = f(\cos \theta), \hspace{10pt} \phi_k = \frac{k \pi}{n}, \hspace{10pt} x_k = \cos(\phi_k).
\end{equation}
\end{linenomath*}
Equation \eqref{eq-discrete-chebyshev-transform} can be rewritten as
\begin{linenomath*}
\begin{equation}\label{eq-dct-intermediate1}
	c_k = \frac{2}{n} \sideset{}{''}\sum_{j=0}^{n} g\left(\frac{j \pi}{n}\right) \cos \left(\frac{j k \pi}{n}\right) \hspace{20pt} (k = 0, 1, \cdots, n).
\end{equation}
\end{linenomath*}
Since $\cos \theta$ and thus $g(\theta)$ are even functions of $\theta$, we can rewrite Equation \eqref{eq-dct-intermediate1}
\begin{linenomath*}
\begin{equation}\label{eq-dct-intermediate2}
	c_k = \frac{1}{n} \sideset{}{''}\sum_{j=-n}^{n} g\left(\frac{j \pi}{n}\right) \exp\left(i \frac{j k \pi}{n}\right) \hspace{20pt} (k = -n, -n + 1, \cdots, n).
\end{equation}
\end{linenomath*}
Furthermore, since $\cos \theta$ and thus $g(\theta)$ are $2 \pi$-periodic functions of $\theta$, we can rewrite Equation \eqref{eq-dct-intermediate2} in the form of discrete Fourier transform
\begin{linenomath*}
\begin{equation}
	c_k = \frac{1}{n} \sum_{j=0}^{2n - 1} g\left(\frac{j \pi}{n}\right) \exp\left(i \frac{j k \pi}{n}\right) \hspace{20pt} (k = 0, 1, \cdots, 2n-1).
\end{equation}
\end{linenomath*}

\section{Evaluation of the Chebyshev sum \label{appendix-clenshaw}}
The Clenshaw algorithm is a recursive method to calculate the sum of Chebyshev polynomials. Let us consider a general sum
\begin{linenomath*}
\begin{equation}\label{eq-cheb-sum}
	S_n(x) = \sum_{j = 0}^{n} a_j P_j(x),
\end{equation}
\end{linenomath*}
where $P_j(x)$ satisfies the recurrence relation
\begin{linenomath*}
\begin{equation}\label{eq-reccurrence-relation}
	P_{r+1}(x) + \alpha_r P_{r}(x) + \beta_r P_{r-1}(x) = 0,
\end{equation}
\end{linenomath*}
and $\alpha_r$, $\beta_r$ may be functions of $x$ as well as of $r$.

We construct the sequence $b_n, b_{n-1}, \cdots, b_0$, where $b_{n+1} = b_{n+2} = 0$ and 
\begin{linenomath*}
\begin{equation}\label{eq-bsequence}
	b_r + \alpha_r b_{r+1} + \beta_{r+1} b_{r+2} = a_r, \hspace{20pt} (r = n, n-1, \cdots, 0).
\end{equation}
\end{linenomath*}
By replacing $a_j$ in Equation \eqref{eq-cheb-sum} with the sequence $\{b_j\}$ and using the recurrence relation \eqref{eq-reccurrence-relation}, we obtain
\begin{linenomath*}
\begin{equation}
	S_n(x) = b_0 P_0(x) + b_1 \{\alpha_0 P_0(x) + P_1(x)\}.
\end{equation}
\end{linenomath*}
In the case of Chebyshev polynomials, we have
\begin{linenomath*}
\begin{equation}
	P_r(x) = T_r(x), \hspace{10pt} \alpha = - 2 x, \hspace{10pt} \beta = 1.
\end{equation}
\end{linenomath*}
The recurrence relation is
\begin{linenomath*}
\begin{equation}
	b_r - 2x b_{r+1} + b_{r+2} = a_r, \hspace{20pt} (r = n, n-1, \cdots, 0).
\end{equation}
\end{linenomath*}
The Chebyshev sum is
\begin{linenomath*}
\begin{equation}
	S_n(x) = \sum_{j = 0}^{n} a_j T_j(x) = b_0 -  b_1 x.
\end{equation}
\end{linenomath*}

\acknowledgments
%Enter acknowledgments, including your data availability statement, here.
We are grateful to J.~Hohl for editing the manuscript. The work was supported by NASA awards 80NSSC18K1122, 80NSSC20K1788, 80NSSC20K0917, and NSF award 2108582. We would like to acknowledge high-performance computing support from Cheyenne (\url{doi:10.5065/D6RX99HX}) provided by NCAR's Computational and Information Systems Laboratory, sponsored by the National Science Foundation.

%% ------------------------------------------------------------------------ %%
%% References and Citations

%%%%%%%%%%%%%%%%%%%%%%%%%%%%%%%%%%%%%%%%%%%%%%%
%
% \bibliography{<name of your .bib file>} don't specify the file extension
%
% don't specify bibliographystyle
%%%%%%%%%%%%%%%%%%%%%%%%%%%%%%%%%%%%%%%%%%%%%%%

\bibliography{sampling}

%Reference citation instructions and examples:
%
% Please use ONLY \cite and \citeA for reference citations.
% \cite for parenthetical references
% ...as shown in recent studies (Simpson et al., 2019)
% \citeA for in-text citations
% ...Simpson et al. (2019) have shown...
%
%
%...as shown by \citeA{jskilby}.
%...as shown by \citeA{lewin76}, \citeA{carson86}, \citeA{bartoldy02}, and \citeA{rinaldi03}.
%...has been shown \cite{jskilbye}.
%...has been shown \cite{lewin76,carson86,bartoldy02,rinaldi03}.
%... \cite <i.e.>[]{lewin76,carson86,bartoldy02,rinaldi03}.
%...has been shown by \cite <e.g.,>[and others]{lewin76}.
%
% apacite uses < > for prenotes and [ ] for postnotes
% DO NOT use other cite commands (e.g., \citet, \citep, \citeyear, \nocite, \citealp, etc.).
%

\end{document}